# Error Model of Radio Fingerprint and PDR Fusion Indoor Localization

**Haojun Ai, Kaifeng Tang, Sheng Zhang and Yuhong Yang**

Key Laboratory of Aerospace Information Security and Trusted Computing, Ministry of Education, China

School of Cyber Science and Engineering, Wuhan University, China

**Abstract:** Multi-source fusion positioning is one of the technical frameworks for obtaining sufficient indoor positioning accuracy. In order to evaluate the effect of multi-source fusion positioning, it is necessary to establish a fusion error model. In this paper, we first use the least squares method to fuse the radio fingerprint and the PDR positioning, and then apply the variance propagation laws to calculate the error distribution of indoor multi-source localization methods. Based on the fusion error model, we developed an indoor positioning simulation system. The system can give a better positioning source layout scheme under a given condition, and can evaluate the signal strength distribution and the error distribution.

1.Introduction

With a increasing demand for Location Based Services (LBS) in both indoor and outdoor environments, significant progress in indoor localization technology was made in recent years. Researchers have proposed various solutions for indoor positioning of smart phones. Most systems are classified into infrastructure-based(e.g., Received Signal Strength (RSS) fingerprinting) and non-infrastructure-based ones (e.g., Pedestrian Dead Reckoning (PDR) ) [1].

The RSS fingerprinting is a kind of the localization method using the received signal strength RSS from the radio beacons, such as WIFI access points [2,3], Bluetooth devices [4], and cellular telephone towers [5]. Although the positioning accuracy with the RF fingerprinting algorithm was improved in the past years, the stability of poisoning results is still poor due to the multipath effect and device heterogeneity [6]. The PDR-based method, also known as an Inertial Navigation System (INS), is a self-contained localization system that relies on inertial sensors, such as accelerometer, gyroscope, and magnetometer [7,8]. Because the PDR-based method in a smartphone relies on an Inertial Measurement Unit (IMU), whose positioning error accumulates over time, it is not suitable for long-term positioning. To achieve better indoor localization results, many smartphone-based localization approaches use various filters, such as the Kalman Filter (KF), Unscented Kalman Filter (UKF) and Particle Filter (PF) to integrate PDR and fingerprint methods [9–11]. Among them the PF is the most popular providing the best localization results. For the localization scheme based on PF, the location of a user is predicted by user motion parameters measured with inertial sensors and then corrected by positional information obtained from radio fingerprints [12–14].

Many indoor LBS applications, require a quick assessment of the radio environment of the area to see whether it can meet the positioning accuracy requirements. In order to realize high-precision and high-availability indoor positioning scheme, it is necessary to establish an accurate indoor positioning error model for different environments. The error model can provide a technical basis for the optimal layout of the positioning infrastructure and explore the factors affecting the positioning accuracy [15-17]. There are many articles about the error model of radio fingerprint and the error model of PDR, but very little focus on the theoretical analysis of positioning accuracy based on a fusion of radio fingerprint and PDR.

The deployment of base stations in the past was mainly based on the experiences which leads to an extreme deployment cost. So we need an indoor positioning system to give a near-optimal positioning source laying solution and the corresponding error distribution[18]. The system requires an error model to evaluate the quality of the laying solution.

In this paper, we present an error model based on the least squares method for radio fingerprint and PDR fusion scheme, analyze the factors affecting positioning accuracy, and evaluate it in a real environment. The evaluation results show that the presented model is effective. We propose a method for calculating the spatial error distribution. Then we develop an indoor positioning system, and the fusion error model is applied, which is used as the basis for the deployment of the positioning sources and help visualize the error distribution.

The rest of the paper is organized as follows. The second section gives a review on the related studies on fingerprint and PDR error models. The third section established error models for radio fingerprint, PDR, and their fusion. The fourth section performs some simulations and experiments in a situation. The fifth section describes the developed indoor positioning simulation system. The last section concludes this research.

**2. Related work**

The errors of radio fingerprint mainly come from multipath effect and device heterogeneity [6]. Multipath effect is the interference of radio signals of more than one paths. When a signal during its propagation encounters obstacles it will reflect, refract, and diffract, resulting in the receiver receiving multipath signals. The device heterogeneity means different devices will receive different signal strengths at the same location and from the same source. At present, the research on the error model of RF fingerprints is mainly based on the Cramer-Rao Lower Bound (CRLB).

Tichavsky et al. [19] provided an expression of the recursive posterior CRLB of a nonlinear filter based on Bayesian framework. Qi [20] proposed a generalized CRLB (called G-CRLB) of a wireless system for NLOS (no line-of-sight) environment. Having analyzed hybrid line-of-sight (LOS)/NLOS environment Qi indicated that with a prior knowledge of wireless transmission channel the performance can be improved. Hossain et al. [16] analyzed the CRLB model with WIFI Signal Strength Difference (SSD) as a fingerprint. When there is a problem of device heterogeneity, SSD is a good choice as fingerprint. Zhou et al. [15] studied the CRLB model of WIFI fingerprints under different signal distribution conditions and mixed signal distribution conditions. Saliha et al. [21] used a log-mixed model, instead of a log-shadow model, to calculate the error bound. Ai et al. [22] proposed a CRLB model related to the window size of RSS. Lei [23] proposed a CRLB model based on LOS and RSS, which reduces the error of distance estimation by distinguishing the LOS and NLOS components in the attenuated signal. Elina [17] extended the CRLB model to the situation in a 3D scene, and studied the impact of the topology of the APs and the deployment density of the APs on the error model. Zhao [24] proposed a model of ER-CRLB, which is an extended recursive CRLB. The model can be adapted to complex and variable environments, considering a priori information and various uncertain factors. Tian et al. [25] proposed a probabilistic model to clarify the indoor positioning performance based on RSS fingerprints, analyzed the probability with which users are located in a certain area, and revealed the interactions among accuracy, reliability and measured valuesduring the positioning process. Li [26] used the method based on the nonlinear least squares to derive the closed expression of the positioning error and obtained a new lower bound smaller than CRLB.

The PDR error is mainly caused by the errors of step frequency detection, step length estimation and heading estimation. Since the PDR itself cannot correct errors, the positioning error will accumulate as the user walks for a longer time.

Lachapelle et al. [27] proposed three step length error models: Gaussian model, constant random model, and Gauss Markov model. Jahn et al. [28] established the error models for four methods of measuring step length, and discussed the systematic and random errors with the Taylor expansion. The first model is based on the biomechanical model of Alvarez [29], measuring acceleration at the user's center of mass. The second model is based on the 4th power between the maximum and minimum values of the step size and the single-step internal acceleration signal proposed by Weinberg [30]. The last two models are the error analysis of the empirical models of Bylemans [31] and Kim [32]. The error in heading estimation is mainly due to magnetic field interference and gyroscope drift. In the early days, Lachapelle et al. [26] modeled the error of the gyroscope as a random constant deviation when establishing the PDR error model, so the heading error was considered to be linear with time. Subsequently, in Chen [33], the predictable errors of the magnetic field, including hard and soft iron effects, magnetic declination, tilt and misalignment, are derived in detail, and a unified heading error model is obtained.

The multi-source fusion localization methods mainly study how to fuse fingerprint and PDR by specific algorithms, such as Particle Filter [34] or Kalman Filter [35], but there has few researches on the theoretical error model after their fusion. Zhuo [36] established an error model for RFID and WIFI fingerprints and obtained a closed-form solution. Tarrío [37] simply derived the error model of RSS and PDR, and analyzed the relationship between positioning accuracy and energy consumption. But they did not explored the variance propagation model, which is useful for evaluating positioning performance. On the basis of the previous researches, this paper combines fingerprint error model and PDR error model by using the least squares method, and obtains a new error model of indoor localization with fingerprint and PDR fusion.

**3. Proposed Error model**

In this section, we first summarize RSS error model [15] and PDR error model [26], then propose the error model of radio fingerprint and PDR fusion, finally simulate the error distribution of PDR integrated with high-precision signal sources.

**3.1 RSS-based localization error**

Usually the log-distance path-loss model is used to establish the connection of the RSS to the distance between two nodes:

$$P(d) = P(d_0) - 10\beta \log(\frac{d}{d_0}) + \eta \qquad (1)$$

where P(d) and P(d$_0$) stand for the RSSs recorded at the locations with distances d and d$_0$ from the AP, respectively. $\beta$ is the pass loss exponent, and η is the measurement error.

Assuming that $\hat{\theta}=(\hat{x},\hat{y})^T$ is the estimated location of coordinates x and y with respect to its true location θ= (x, y)$^T$, then its covariance matrix is

$$\text{Cov}_\theta(\hat{\theta})=E\{(\hat{\theta}-\theta)(\hat{\theta}-\theta)^T\}=\begin{bmatrix}\sigma_{\hat{x}}^2 & \sigma_{\hat{x}\hat{y}}^2 \\ \sigma_{\hat{y}\hat{x}}^2 & \sigma_{\hat{y}}^2\end{bmatrix} \quad (2)$$

The diagonal elements of (2) represent the variances, and the off-diagonal elements are the covariance. According to the definition of the CRLB, it is the inverse of the Fisher Information Matrix (FIM):

$$\text{Cov}_\theta(\hat{\theta}) \geq J(\theta)^{-1} \quad (3)$$

The FIM, J(θ), is the variance of this score function:

$$J(\theta)=E[(\frac{\partial \ln f(P;\theta)}{\partial \theta})^2]=-E[\frac{\partial^2 \ln f(P;\theta)}{\partial \theta^2}] \quad (4)$$

where f(P; θ) denotes the probability density function of observations $P$ at point $\theta$. The score function is defined as the gradient of its log-likelihood. As the RSSs follow the Gaussian signal distribution, the joint probability density function (PDF) of the RSSs from $\theta$ is calculated as:

$$f(P;\theta)=\prod_{k=1}^{m}\frac{1}{\sqrt{2\pi}\sigma}\exp\left(-\frac{\varepsilon^2}{2\sigma^2}\right) \quad (5)$$

where $\varepsilon$ = P(d) - P(d$_0$) - 10$\beta$log($\frac{d}{d0}$), d=$\sqrt{(x-x_k)^2+(y-y_k)^2}$, $(x_k, y_k)$ represents the coordinates of the *k*-th AP, *m* is the number of APs, and $\sigma^2$ is the variance of the RSSs collected by the receiver. In addition, J(θ) can be written as

$$J(\theta)=\begin{bmatrix}J_{xx}(\theta) & J_{xy}(\theta) \\ J_{yx}(\theta) & J_{yy}(\theta)\end{bmatrix} \quad (6)$$

Combining (4), (5), and (6) gives:

$$J_{xx}(\theta)=(\frac{10\beta}{\sigma \ln 10})^2\sum_{k=1}^{m}[\frac{\cos\alpha}{d}]^2 \quad (7)$$

$$J_{xy}(\theta)=J_{yx}(\theta)=(\frac{10\beta}{\sigma \ln 10})^2\sum_{k=1}^{m}\frac{\sin\alpha\cos\alpha}{d^2} \quad (8)$$

$$J_{yy}(\theta)=(\frac{10\beta}{\sigma \ln 10})^2\sum_{k=1}^{m}[\frac{\sin\alpha}{d}]^2 \quad (9)$$

$\alpha$ is the angle between the line of the AP and the user and the ground, and $d$ is the distance between the AP and the user. Therefore, we can calculate J(θ)$^{-1}$ with

$$J(\theta)^{-1}=\frac{1}{|J(\theta)|}\begin{bmatrix}J_{yy}(\theta) & -J_{yx}(\theta) \\ -J_{xy}(\theta) & J_{xx}(\theta)\end{bmatrix} \quad (10)$$

Taking (7), (8), (9), and (10) into (3), the CRLB for the localization estimation can be expressed as

$$\text{var}(\hat{\theta})=\sigma_{\hat{x}}^2+\sigma_{\hat{y}}^2 \geq \frac{J_{yy}(\theta)}{|J(\theta)|}+\frac{J_{xx}(\theta)}{|J(\theta)|}=\frac{J_{xx}(\theta)+J_{yy}(\theta)}{J_{xx}(\theta)*J_{yy}(\theta)-J_{xy}(\theta)*J_{yx}(\theta)} \quad (11)$$

**3.2 PDR localization error**

As known, PDR estimates the trajectory of an object by continuously adding its displacement from a given starting point. The PDR error is therefore mainly caused by step frequency detection, step length estimation and heading estimation. Integration of the errors of accelerometers and gyroscopes leads to cumulative errors. Because PDR itself cannot correct the error, the positioning error of will continue to accumulate as the user walks for a longer time. The accumulation of the error is also called drift, which exists in all relative positioning systems.

Figure 1 shows the error range of the PDR. A pedestrian starts from point S and proceeds in the AB direction. The point O is the actual position after the N steps. Let $\sigma_S$ be the standard deviation of the distance error, in the walking direction, and $\sigma_G$ be the standard deviation of the distance error in the perpendicular direction.

Point A and point B indicate the error range in the straight direction due to the step error and the heading error. The calculation formula of $\sigma_S$ is as follows:

$$\sigma_S = \sum_{K=1}^{N} S_k \cdot [1 - \cos(\int_{t_1}^{t_k} D^{max} dt)] + \sigma_{SN} \qquad (12)$$

where N is the number of steps, $S_k$ is the step length of step k, $D^{max}$ is the maximum drift speed of the heading, the unit is rad/sec, and $\sigma_{SN}$ is one of the three step length error models. Here the Gaussian model is selected.

Points A⁻, O⁻, and B⁻ represent the maximum negative drift of heading due to gyroscope drift. Points A⁺, O⁺, B⁺ represent the maximum forward drift of the heading due to gyroscope drift. The formula for calculating $\sigma_G$ is as follows:

$$\sigma_G = \sum_{K=1}^{N} S_k \cdot \sin(\int_{t_1}^{t_k} D^{max} dt) \qquad (13)$$

The shaded area in the figure is the PDR error range after the pedestrian walks N steps, which can be approximated by the Root Mean Square Error (RMSE):

$$\text{RMSE} = \sqrt{\sigma_S^2 + \sigma_G^2} \qquad (14)$$

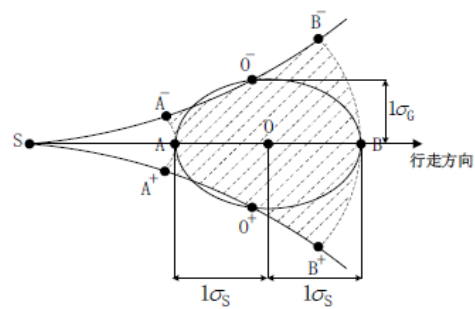

Figure.1 Error model of PDR

### 3.3 Fused localization error

The positioning accuracy of an indoor positioning system depends on the number of observations and their accuracies. Under the same conditions, more observations, more accurate the results will be. Compared with a single-sensor positioning system, the multi-sensor architecture will provide more accurate and reliable results. To best fuse the data of different sensors the least squares principle is employed. The main idea of the least squares method is to use all the relevant observations to solve for the unknown parameters in such a way that the sum of the squared differences between the theoretical values and the observations (the error, or the residual) is minimized. Let the observation equations of n sensors for a certain performance parameter be:

$$Y = Hx + e \qquad (15)$$

Where x is the parameter to be estimated, Y is the vector of n measurement, e is the noise vector, and H is a n by 1 matrix. The least squares estimation of x is to minimize the sum of squares of the weighted errors

$$J_w(\hat{x}) = (Y - H\hat{x})W(Y - H\hat{x}) \qquad (16)$$

where W is a positive definite diagonal weight matrix, W=diag($w_1, w_2 ... w_n$). Find the partial derivatives of $J_w(\hat{x})$ with respect to x and make them equal to 0. The least squares estimate of the estimator $\hat{x}$ is

$$\hat{x} = (H^T W H)^{-1} H^T W Y = \frac{\sum_{i=1}^{n} w_i y_i}{\sum_{i=1}^{n} w_i} \qquad (17)$$

Assuming the multiple independent random variables follow normal distribution, the measurement noise is also normal and the variance for i-th sensor reads

$$E[e_i^2] = E((x - y_i)^2) = \sigma_i^2 \qquad (18)$$

The variance of the estimated parameter x is

$$E[(x - \hat{x})^2] = E[(x - \frac{\sum_{i=1}^{n} w_i y_i}{\sum_{i=1}^{n} w_i})^2]$$

$$= E[\sum_{i=1}^{n} (\frac{w_i}{\sum_{i=1}^{n} w_i})^2 (x - y_i)^2]$$

$$= \sum_{i=1}^{n} (\frac{w_i}{\sum_{i=1}^{n} w_i})^2 \sigma_i^2 \qquad (19)$$

The weight of an observation is normally defined as

$$w_i = \frac{1}{\sigma_i^2} \quad i = 1,2, ... n \qquad (20)$$

Substituting (20) in to (19), one get the variance of the estimated parameter derived from the two sensor data

$$\sigma^2 = E[(x - \hat{x})^2] = \frac{\sigma_1^2 \sigma_2^2}{\sigma_1^2 + \sigma_2^2} \qquad (21)$$

In a system, the RSS-based localization is done periodically and inertial sensors are used between the consecutive RSS-localizations to track the position of the user. The total localization error will have two components, one from the RSS-localization and the other from the PDR localization:

$$RMSE^2(t) = f(RMSE_{RSS}^2, RMSE_{PDR}^2(t)) \qquad (22)$$

where *t* is the elapsed time from the last localization reset, and RMSE is Root Mean Square Error. The RSS-localization error does not depend on the time *t*, but on the positions of the nodes and the characteristics of the channel. The PDR localization error depends on the integration time *t*, step length, and heading. The maximum localization error appears at the end of the integration time.

Assume that the position $\theta_1 = (x_1, y_1)$ derived from the RSS fingerprint, the variances of the coordinates x and y are ($\sigma_{x1}^2$, $\sigma_{y1}^2$), and the position $\theta_2 = (x_2, y_2)$ derived from the PDR, the variance of the coordinates x and y are ($\sigma_{x2}^2$, $\sigma_{y2}^2$). We use a fusion method based on the least squares method. The position after the fusion is $\theta = (x_R, y_R)$, and the variances of the coordinates x and y are ($\sigma_{x_R}^2$, $\sigma_{y_R}^2$). Taking them into the above (17) (21) results in:

$$x_R = \frac{\sum_{i=1}^n w_i x_i}{\sum_{i=1}^n w_i} = \frac{x_1 \cdot \frac{1}{\sigma_{x1}^2} + x_2 \cdot \frac{1}{\sigma_{x2}^2}}{\frac{1}{\sigma_{x1}^2} + \frac{1}{\sigma_{x2}^2}} = \frac{x_1 \cdot \sigma_{x2}^2 + x_2 \cdot \sigma_{x1}^2}{\sigma_{x1}^2 + \sigma_{x2}^2}$$

$$= \begin{bmatrix} \frac{\sigma_{x2}^2}{\sigma_{x1}^2+\sigma_{x2}^2} & \frac{\sigma_{x1}^2}{\sigma_{x1}^2+\sigma_{x2}^2} \end{bmatrix} \begin{bmatrix} x_1 \\ x_2 \end{bmatrix} \qquad (23)$$

$$y_R = \begin{bmatrix} \frac{\sigma_{x2}^2}{\sigma_{x1}^2+\sigma_{x2}^2} & \frac{\sigma_{x1}^2}{\sigma_{x1}^2+\sigma_{x2}^2} \end{bmatrix} \begin{bmatrix} y_1 \\ y_2 \end{bmatrix} \qquad (24)$$

$$\sigma_{x_R}^2 = \begin{bmatrix} \frac{\sigma_{x2}^2}{\sigma_{x1}^2+\sigma_{x2}^2} & \frac{\sigma_{x1}^2}{\sigma_{x1}^2+\sigma_{x2}^2} \end{bmatrix} \begin{bmatrix} \sigma_{x1}^2 & \sigma_{ab} \\ \sigma_{ba} & \sigma_{x2}^2 \end{bmatrix} \begin{bmatrix} \frac{\sigma_{x2}^2}{\sigma_{x1}^2+\sigma_{x2}^2} \\ \frac{\sigma_{x1}^2}{\sigma_{x1}^2+\sigma_{x2}^2} \end{bmatrix}$$

$$= [\frac{1}{\sigma_{x1}^2+\sigma_{x2}^2}]^2 [\sigma_{x2}^2, \sigma_{x1}^2] \begin{bmatrix} \sigma_{x1}^2 & 0 \\ 0 & \sigma_{x2}^2 \end{bmatrix} \begin{bmatrix} \sigma_{x2}^2 \\ \sigma_{x1}^2 \end{bmatrix}$$

$$= \frac{\sigma_{x1}^2 \sigma_{x2}^2}{\sigma_{x1}^2+\sigma_{x2}^2} \qquad (25)$$

$$\sigma_{y_R}^2 = \frac{\sigma_{y1}^2 \sigma_{y2}^2}{\sigma_{y1}^2+\sigma_{y2}^2} \qquad (26)$$

Referring to sections 3.1and 3.2, $\sigma_{x1}^2$ is $\sigma_x^2$ in the RSS-based error model, $\sigma_{x2}^2$ is $\sigma_S^2$ in the PDR error model, $\sigma_{y1}^2$ is $\sigma_y^2$ in the RSS-based error model, and $\sigma_{y2}^2$ is $\sigma_G^2$ in the PDR error model. Considering all these the root mean square error of the estimated position with RSS fingerprint and PDR fusion:

$$RMSE = \sqrt{\sigma_{x_R}^2 + \sigma_{y_R}^2} = \sqrt{\frac{\sigma_x^2 \sigma_S^2}{\sigma_x^2+\sigma_S^2} + \frac{\sigma_y^2 \sigma_G^2}{\sigma_y^2+\sigma_G^2}} \qquad (27)$$

It can be seen from the above equation that the error after data fusion is not greater than the errors of the two individual methods. The root mean square error is related to the position of the node, the characteristics of the channel, the time t, the step length and the heading. Using this model, the error after the fusion can be estimated, and the error distribution in the space can be calculated.

## 4. Simulation and Experiment

### 4.1 Experimental environment and data collection

The experimental environment is in an underground parking lot with a length of about 102 meters and a width of about 33 meters. The area is nearly 3,400 square meters. We arranged 106 iBeacons at intervals of 4 to 5 meters , as shown in Figure 4. The left side is the parking lot exit, and the right side is the parking lot entrance. A pedestrian held a nexus5 mobile phone and went 50m from start point to collect iBeacon Mac address, RSSI, signal data of accelerometer and gyroscope during the walking.

In order to calculate the relationship between the real-time positioning error and the number of walking steps N, it is necessary to record the true position of each step of the pedestrian during walking. However, since the GPS signal is unreachable in the indoor environment, so this experiment uses the devices shown in Figure 5 below to record the real position of the pedestrian. Among them, the No. 1 device is a Nexus5 mobile phone held by pedestrians during walking, the No. 2 and No. 3 devices are respectively a measuring tape with a total length of 50 meters and a label attached to the measuring tape for amplifying the meter scale value, the No. 4 device is a high-definition Canon camera that records the user's real-time position on the tape measure.

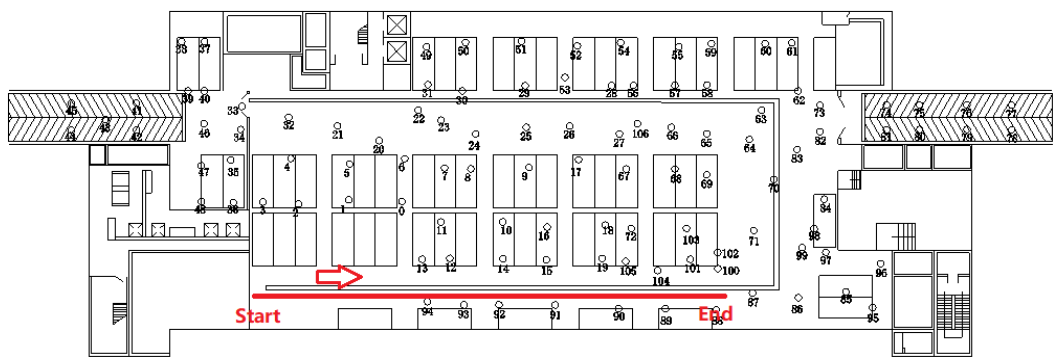

Figure.4 Experimental environment

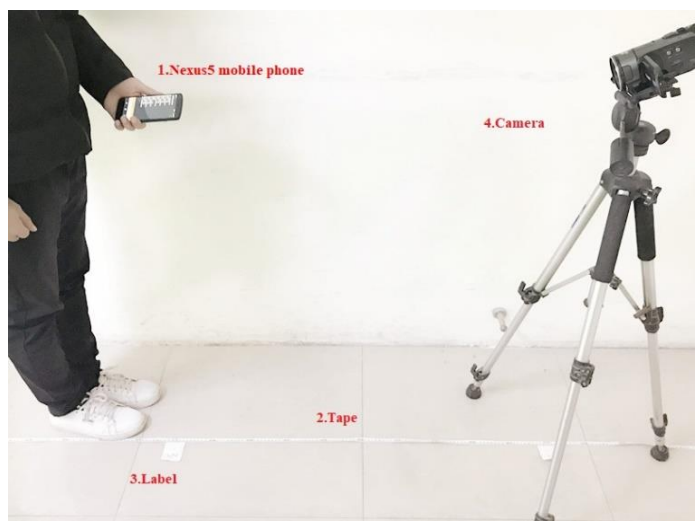

Figure.5 Experimental devices

**4.2 Comparison of three error models**

In the above environment, we first simulated the three error models of RSS (Equation 11), PDR (Equation 14) and the fusion of the two (Equation 27). The parameters used in the simulation are listed in Table 1 and the results are shown in Figure 6. It can be seen that the error model of the Bluetooth fingerprint is related to the position of the beacon on the walking path. Several obvious turning points in the figure are at the time when the pedestrian walks directly under the beacon. At these points the accuracy of Bluetooth positioning is higher. The error in PDR is accumulating over time, as expected. The error after data fusion is in the beginning smaller, close to the error curve of PDR, due to the smaller error of PDR. As the number of steps increases, the error of PDR increases, and the fusion result is similar to the error curve of Bluetooth fingerprint. As we can see, the fluctuation after fusion is smaller than before, and the error after fusion is also smaller than the errors of two separate positioning methods, so the fusion of Bluetooth and PDR can effectively reduce the positioning error.

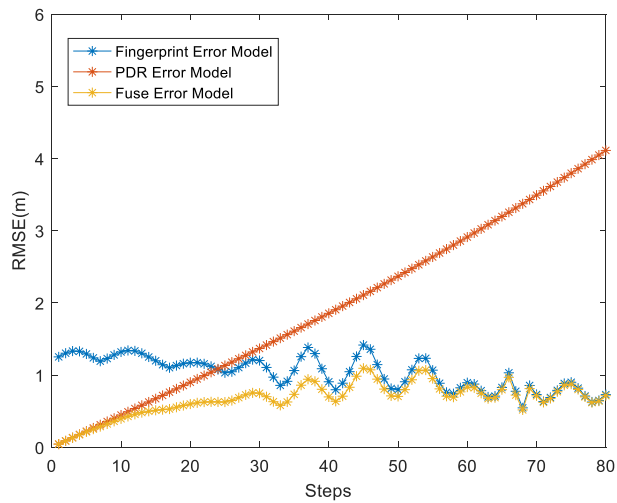

Figure.6 Simulation results of three error models

Table 1. Parameters used in the simulation

| RSS error model | $\beta$=3, $\sigma$=1.732dB$_m$ |
|---|---|
| PDR error model | $S_k$=0.625m, $D^{max}$=0.0283 rad/sec, $\sigma_{SN}$=0.0446m |

**4.3 Comparison of fusion error model and experimental results**

In this section we compare the experimental results in the real situation with our fusion error model to prove the theoretical analysis. In order to ensure the reliability of the experiment, a total of five tests were conducted. We used the particle filter to fuse the fingerprint information with the PDR data. The experiment results are calculated on the Matlab2016 platform, and shown in Figure 7, five straight lines are the result of the experiments and blue star line is the fusion error model.

It can be seen from Figure 7 that due to various uncertain factors in the tests, the positioning errors fluctuate somewhat, but the trend is similar, the result of test 1 is very close to the model, which shows that the proposed fusion error model can reflect the error distribution and trend after fingerprint and PDR fusion. Figure 8 shows the 2D simulation results of the spatial error distribution, which are consistent with the fluctuation of the curve in Figure 7.

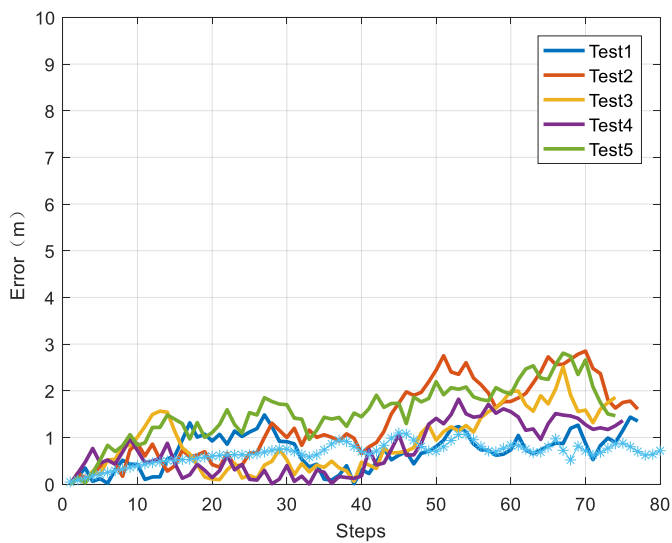

Figure.7 Experimental results

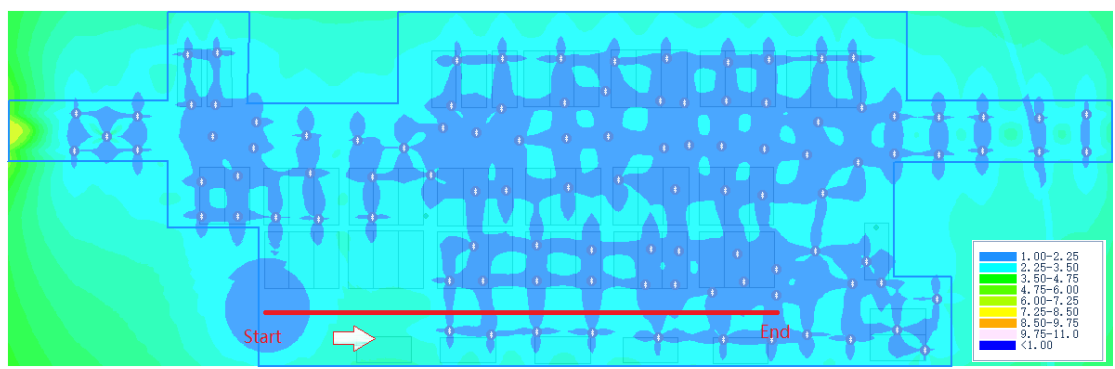

Figure.8 2D simulation results

## 5. Development of Indoor Positioning Simulation System (IPS)

In this section, we will apply the fusion error model introduced in this paper to an indoor positioning simulation system we develop, it can be used to evaluate the quality of the signal source deployment and visualize the error distribution after the PDR method is integrated.

### 5.1 System description

With the wide applications of indoor positioning services, users have higher and higher requirements for positioning accuracy. The deployment of base stations in the past was mainly based on the experiences, and no guidelines exist. Therefore, we developed a new indoor positioning simulation system. It can provide a near-optimal signal source deployment in a given indoor environment. The system can visualize signal strength and positioning accuracy. It will improve work efficiency and layout quality, and achieve high-precision positioning in different indoor environments. The system is developed with C# language in Windows 10, and the platform Visual Studio 2015 and the ArcGIS Engine are used.

### 5.2 System structure

Figure 9 shows the flow chart of the developed indoor positioning simulation system.

First, you need to load the map file, and then set the layout conditions. Here you can select different types and quantities of signal sources. You also need to select the automatic layout algorithm. These are used as inputs, then the system does calculation using the automatic layout algorithm under given conditions and provides the optimal layout. After the simulation is completed, you can view the simulation results, including the signal strength map, positioning error distribution map and other simulation results. If the user is not satisfied with the simulation result, the number and location of the signal sources can be adjusted in the user interaction interface.

Figure 10 is a screenshot of the interface. It can be seen that the upper column is the menu and toolbox, the left column is the layer management, the right column is the project management, and the middle is the indoor map.

**5.3 Simulation results**

We applied the developed positioning simulation system to an office building, which is about 94 meters long and 39 meters wide with a corridor in the middle and a hall below. We selected 15 iBeacons as the signal source. The propagation model uses a log-distance path-loss model, and the layout algorithm uses a simulated annealing algorithm.

Figure 11 shows the signal strength map after the simulation. The signal strength map reflects the distribution of signal strength in the whole space. The user can judge whether the signal source is arranged reasonably from the signal strength map, if there are some places where the signal strength is weak, the signal sources can be added or moved using the user interaction, and the signal strength map can be dynamically changed as the positioning source increases or moves.

Figure 12 shows the positioning error map after the completion of the layout in the case of only Bluetooth. The average error of the whole indoor environment is 3.2 meters. However, due to the characteristics of Bluetooth, the accuracy is better only in the area close to a signal source. As shown in the figure, most of the area is in the colors of light blue and green, which means the error is between 2.25m and 4.75m. The area of the error below 2m is within a circle around the beacon.

Figure 13 is the positioning error map after the PDR is included. We apply the fusion error model in this paper to the simulation system. It can be seen that after the PDR is fused, the average error in the whole indoor environment is reduced to 1.5 meters, which is almost covered by the blue area. This is close to the actual positioning effect. Therefore, PDR can improve the positioning accuracy in the areas between high-precision positioning sources.

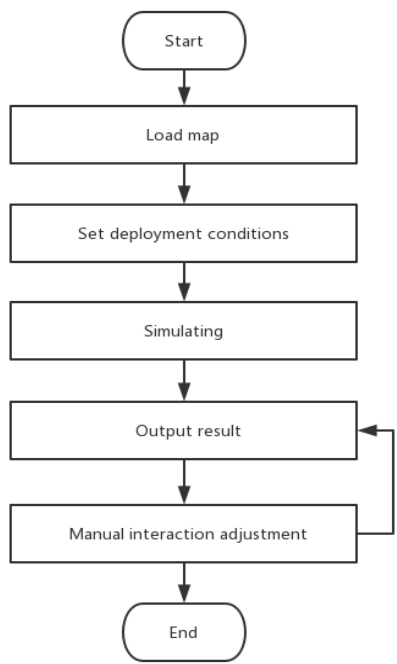

Figure.9 Simulation system flow chart

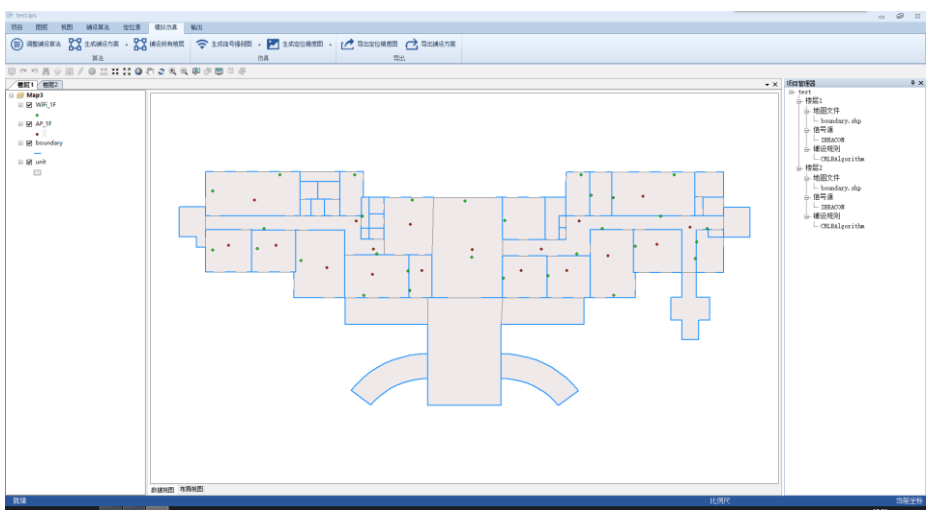

Figure.10 Simulation system interface

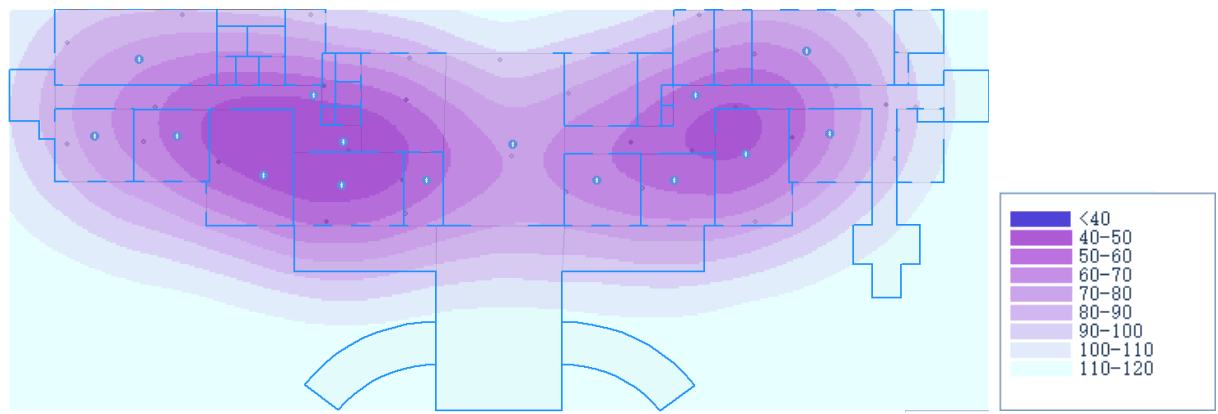

Figure.11 Signal strength map

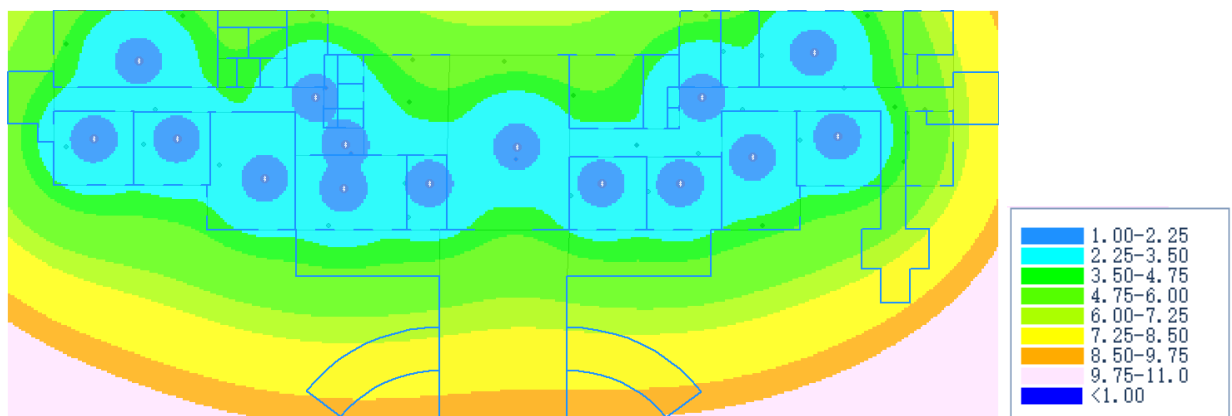

Figure.12 Positioning error distribution map

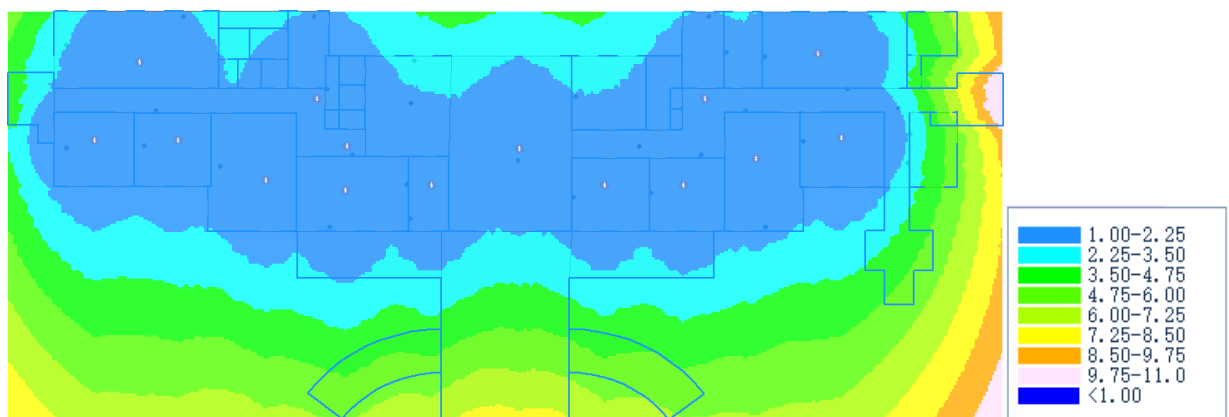

Figure.13 Positioning error distribution map with PDR

## 6. Conclusion

In this paper, we first analyze and summarize the classical error models of RSS fingerprint and PDR. On these basses we proposed a error model of fusing RSS and PDR. This fusion error model is compared with the results in the actual situation, to proves that the model is effective. We also proposes a method for calculating the error distribution in space. To facilitate practical applications we developed an indoor positioning simulation system, which applies our fingerprint positioning and PDR fusion error model. The system can not only complete the automatic deployment of signal sources, but also visualize the signal strength distribution and error distribution.

In the future, we plan 1) consider the map information of the room structure, walls, ceilings, etc. into our error model; 2) continue to study the error model under the condition of multi-source fusion. This paper only studies the error model of fingerprint localization and PDR fusion. Later, various signal sources, such as sound, light, magnetic and visual can be fused to study their fusion error models, explore factors that affect the accuracy of multi-source fusion positioning.